\documentclass[10pt,conference]{IEEEtran} 

\usepackage{flushend}
\usepackage{cite}
\usepackage{amsmath,amssymb,amsfonts}
\usepackage{algorithmic}
\usepackage{graphicx}
\usepackage{textcomp}
\usepackage{hyperref}
\usepackage{xcolor}
\usepackage{color}
\usepackage{booktabs}
\usepackage{xspace}
\usepackage{graphicx}
\usepackage{multirow}
\usepackage[framemethod=TikZ]{mdframed}
\usepackage{textcomp}
\usepackage{booktabs}
\usepackage{colortbl}
\usepackage{ragged2e}
\usepackage{courier}
\usepackage{textcomp} % For \texttildelow
\usepackage{listings}
\usepackage{varwidth}

\def\BibTeX{{\rm B\kern-.05em{\sc i\kern-.025em b}\kern-.08em
    T\kern-.1667em\lower.7ex\hbox{E}\kern-.125emX}}

\usepackage[normalem]{ulem}

\newcommand{\myparagraph}[1]{\vspace{0.1cm}\noindent \xspace\textbf{\textit{#1.}}}

\newcommand{\eg}{e.g.,\xspace}

\newcommand{\etal}{et al.\xspace}

\newboolean{showcomments}
\setboolean{showcomments}{true}
\ifthenelse{\boolean{showcomments}}
	{\newcommand{\nb}[3]{
		{\colorbox{#2}{\bfseries\sffamily\scriptsize\textcolor{white}{#1}}}
		{\textcolor{#2}{\sf\small$\blacktriangleright$\textit{#3}$\blacktriangleleft$}}}
	 }
	{\newcommand{\nb}[3]{}
	 }

\usepackage{xcolor}
\usepackage{textcomp}
\usepackage[T1]{fontenc}
\usepackage{inconsolata}

\newcommand{\ct}{\lstinline[breakatwhitespace=true, breaklines=true, basicstyle=\footnotesize]}

\newcommand{\nadel}[1]{}  % suggested removal: does not produce anything

\definecolor{pblue}{rgb}{0.13,0.13,1}
\definecolor{pgreen}{rgb}{0,0.5,0}
\definecolor{pred}{rgb}{0.9,0,0}
\definecolor{pgrey}{rgb}{0.46,0.45,0.48}
\definecolor{lightgreen}{rgb}{0.56, 0.93, 0.56}
\definecolor{lightgray}{rgb}{0.83, 0.83, 0.83}

\definecolor{dkgreen}{rgb}{0,0.6,0}
\definecolor{gray}{rgb}{0.5,0.5,0.5}
\definecolor{mauve}{rgb}{0.58,0,0.82}
\definecolor{light-gray}{gray}{0.97}  
\definecolor{weborange}{rgb}{255,165,0}
\definecolor{darkviolet}{rgb}{0.5,0,0.4}
\definecolor{darkgreen}{rgb}{0,0.4,0.2} 
\definecolor{darkblue}{rgb}{0.1,0.1,0.9}
\definecolor{darkgrey}{rgb}{0.5,0.5,0.5}
\definecolor{lightblue}{rgb}{0.4,0.4,1}

\definecolor{stringColor}{rgb}{0.16,0.00,1.00}
\definecolor{fieldColor}{rgb}{0.16,0.00,1.00}
\definecolor{annotationColor}{rgb}{0.39,0.39,0.39}
\definecolor{keywordColor}{rgb}{0.50,0.00,0.33}
\definecolor{commentColor}{rgb}{0.25,0.50,0.37}
\definecolor{javadocColor}{rgb}{0.25,0.37,0.75}
\definecolor{jTagColor}{rgb}{0.50,0.62,0.75}
\definecolor{eTagColor}{rgb}{0.50,0.62,0.75}
\definecolor{lineNumberColor}{rgb}{0.47,0.47,0.47}
\definecolor{shadecolor}{rgb}{0.9,0.9,0.9}

\newcommand\coderef{Figure~\ref}
\newcounter{rmd}

\newcounter{bmd}

\begin{document}

\title{Automatically Generating Single-Responsibility Unit Tests}

\author{\IEEEauthorblockN{Geraldine Galindo-Gutierrez} 
\IEEEauthorblockA{Centro de Investigaci\'on en Ciencias Exactas e Ingenier\'ias - CICEI \\ Universidad Cat\'olica Boliviana \\
widni.galindo@ucb.edu.bo}
}

\twocolumn[
\begin{@twocolumnfalse}
\maketitle
\begin{center}
    \vspace{-1em}
    \small\textit{This paper has been accepted for publication at the 47th International Conference on Software Engineering (ICSE 2025). The final version will appear in the official conference proceedings published by ACM/IEEE.}
    \vspace{1em}
\end{center}
\end{@twocolumnfalse}
]

\begin{abstract}
Automatic test generation aims to save developers time and effort by producing test suites with reasonably high coverage and fault detection. However, the focus of search-based generation tools in maximizing coverage leaves other properties, such as test quality, coincidental. The evidence shows that developers remain skeptical of using generated tests as they face understandability challenges. Generated tests do not follow a defined structure while evolving, which can result in tests that contain method calls to improve coverage but lack a clear relation to the generated assertions. In my doctoral research, I aim to investigate the effects of providing a pre-process structure to the generated tests, based on the single-responsibility principle to favor the identification of the focal method under test. To achieve this, we propose to implement different test representations for evolution and evaluate their impact on coverage, fault detection, and understandability. We hypothesize that improving the structure of generated tests will report positive effects on the tests' understandability without significantly affecting the effectiveness. We aim to conduct a quantitative analysis of this proposed approach as well as a developer evaluation of the understandability of these tests. 
\end{abstract}

\IEEEpeerreviewmaketitle

\section{Problem Statement}
%\del{The AAA pattern, short for \textit{Arrange, Act, Assert} is a widely recommended practice in test writing due to its effects on debugging and maintenance~\cite{khorikov2020unit}. The pattern structures the test in the setup, the focal method, and the verification steps. When following these pattern tests maintain the single responsibility principle and reusable code. This benefits the developer by improving understanding of the relationship between test and source code. Consider, as an example, manually written \ct{testDepositToAccount} in \coderef{fig:manual-test}. If the test fails, the \ct{deposit} method must be reviewed, identified in the \textit{Act} step (line 3). }

Unit tests act as the first line of prevention against potential bugs introduced during software evolution. When a unit test fails, it is required that developers fully understand the behavior under test before any fix. When reading a unit test, developers expect to easily detect the faulty behavior and pinpoint the issue. For this reason, several principles and good practices are suggested for creating readable, maintainable, and reliable tests~\cite{khorikov2020unit,osherove2013art}. For example, consider \ct{testDepositToAccount} in \coderef{fig:manual-test} which contains a descriptive name (a readability good practice) and the main method under test is clearly stated (single-responsibility principle).

\begin{lstlisting}[language = Java, escapechar=^, caption = Test for deposit method in BankAccount class. , label={fig:manual-test}, ]
public void testDepositToAccount() {
  BankAccount account = new BankAccount("X", 100.00);
  account.^\textbf{deposit(50.00)}^;
  assertEquals(150.00, account.getBalance(),0.01);
}
\end{lstlisting}

Despite its advantages, test writing is a complex and time-consuming process for developers~\cite{aniche2022developers,straubinger2023}. Automatic test generation tools address this issue by producing test suites with optimized coverage and limited human intervention~\cite{Fraser2012,Pacheco2007,tonella2004,lukasczyk2022pynguin}. However, their focus on coverage and mutation score leaves other properties, such as readability and maintainability, coincidental~\cite{grano2020pizza,grano2018empirical,lin2019quality,brandt2022}. Studies show that generated tests present quality pitfalls~\cite{panichella2022smells,icsme2023,setiani2022understandable} and Shamshiri \etal found that developers are skeptical of using generated tests due to their unclear purpose~\cite{shamshiri2015automatically}. 
%\del{Brandt \etal and Abdi \etal received understandability observations when they sent generated tests in pull requests~\cite{brandt2024shaken,abdi2022small}}

To exemplify this issue, consider \ct{test17} in \coderef{fig:gen-test}. This test invokes three methods from \ct{BankAccount} class: \ct{closeAccount}, \ct{deposit}, and \ct{transferFunds}; and verifies their effects by checking the account balance. However, it is unclear which of these is the focal method, the behavior under test, as all method calls can be separately verified by the same assertion in line 6. In case of failure, it is not easy to determine which method is not working as expected as the test seems to be checking various methods, thereby violating the single-responsibility principle.

\begin{lstlisting}[language = Java, escapechar=^, caption = Generated test for BankAccount class. , label={fig:gen-test}, ]
public void test17()  throws Throwable  {
  BankAccount bankAccount0 = new BankAccount("", 0.0);
  bankAccount0.^\textbf{closeAccount}^();
  bankAccount0.^\textbf{deposit}^(665.49);
  bankAccount0.^\textbf{transferFunds}^(bankAccount0, 0.05);
  assertEquals(665.49,bankAccount0.getBalance(),0.01);
}
\end{lstlisting}

Different test generation tools fall short to follow the single-responsibility principle for a non-trivial portion of their tests. In consequence, developers have to spend more time understanding the purpose of generated tests before using them. Various approaches have previously attempted to address this problem using post-processing approaches such as minimizing tests to avoid redundancies~\cite{Fraser2011}, improving readability~\cite{Daka2015,daka2017generating,deljouyi2023generating}, or adding test documentation~\cite{roy2020deeptc}. Other approaches improve test quality during search, for example, integrating metrics to avoid test smells during generation~\cite{afonso2023smellfree} or using readability metrics and models~\cite{Daka2015}. However, these approaches are inherently limited by the current functioning of generation tools, where tests are mainly optimized for coverage and do not consider a defined structure. In consequence, the generated test code is often perceived as machine generated~\cite{grano2018empirical,Daka2015,tufano2022methods2test}, which influences the adoption of test generation tools in industry. 

%Different studies with developers recommend improving the understandability and structure of generated tests, reducing the skepticism of developers when using them~\cite{brandt2022,lin2019quality,grano2018empirical}.

\section{Proposed Approach}
Unlike previous approaches, we propose to address this problem by changing the underlying foundations of test generation to produce tests that follow the single responsibility principle by construction in order to improve their understandability for developers.
%hoping to increase their acceptance by practitioners.

The common representation of a test $T$ in state-of-the-art search-based generation tools is a list of statements $[s_1, \ldots, s_n]$ of variable length $n$ where each $s$ has a defined statement type (primitive, constructor, field, method, assignment)~\cite{Fraser2011,lukasczyk2022pynguin}. The $n$ statements participate in crossover and mutation operations focusing on optimizing a coverage-guided fitness function. After reaching 100\% coverage or exhausting the search budget, the tools generate assertions and apply different post-processing steps~\cite{afonso2023smellfree}. However, not considering structural properties may cause multiple methods invoked in the same test increasing coverage, but hindering the purpose of the test.  

\myparagraph{Test Representation using Focal Methods} 
We propose to provide test cases with a defined structure that indicates the focal method under test from construction. Previous studies have used focal methods when generating tests using LLM-based approaches~\cite{tufano2021unit,tufano2022methods2test,he2024empirical}, identifying the focal method in the input. However, there are several considerations to adapt this approach in search-based tools, we must evaluate different structures and their effects in mutation and crossover operations. For example, a possible representation could be to fix the last statement of the test, $s_n$, as the focal method and consider only the $n-1$ statements for crossover and mutation operations. To follow single-responsibility principle, the proposed structures are focused on identifying the statements dedicated to the test setup (\eg object constructor and method invocations) and the statements that contain the tested behavior.
%\ins{Following single-responsibility principle demands to have a sequence of method calls covering a specific scenario, one unit of work~\cite{osherove2013art}. Therefore, our proposed structures are focused on maintaining coherence between method invocations.}

%We propose to change the representation of the test to clearly separate the \textit{Arrange} and \textit{Act}, ensuring the \textit{Act} step consists of focal methods, which are the primarily target for verification. Previous studies have used focal methods when generating tests using LLM-based approaches~\cite{tufano2021unit,tufano2022methods2test,he2024empirical}, identifying the focal method in the input. We plan to use a similar approach and evaluate different test case representations where the focal method is identified before applying generation algorithms. 

 %However, the effects of the representations on the crossover, mutation, and fitness function must be evaluated. 

\myparagraph{Assertion Generation for Single-Responsibility Tests} 
In an initial study, we observed that in a portion of generated tests, the assertions are not clearly related to the invoked methods of the class under test~\cite{icsme2023}. Assertion generation is currently based on mutation analysis, where assertions that uniquely detect at least one new mutant are kept as part of the generated test. However, not all newly detected mutants are necessarily related to the main behavior under test. We propose to adapt assertion generation to our proposed structures and maintain only assertions related to the focal method of the test. With this approach, we aim at grouping assertions based on the method their detected mutants are related to. This process can allow new refactorizations~\cite{panichella2022smells} between groups of tests, which we aim to evaluate and discuss.

% Besides clearly identify the \textit{Arrange} and \textit{Act} steps, the relation between the steps \textit{Act} and \textit{Assert} reflected in the assertion generation process must be coherent. In an initial study, we observed that the assertions are not related to every method of the class under test called~\cite{icsme2023}. We aim to study the effect of our proposed representations on the relation between the focal methods and the assertions generated. Assertion generation is currently based on mutation analysis, where assertions that uniquely detect at least one new mutant are kept as part of the generated test. However, generated assertions may be related to any statement in the generated test and not necessarily to the focal methods.

% The single-responsibility principle should be reflected in the assertions.  

\section{Research Questions}
We hypothesize that using a structure that maintains the single-responsibility principle will improve the quality and understandability of the test code without significantly affecting the effectiveness in terms of coverage and mutation score. We guide our research and the evaluation of our proposal with the following questions.

\textbf{RQ1 - Effectiveness: }\emph{How well do the proposed structures perform in coverage and mutation score compared to the current representation of unit tests?} 

\textbf{RQ2 - Single-Responsibility: }\emph{How effective is the proposed approach in achieving single-responsibility in automatically generated unit tests?}

\textbf{RQ3 - Coherence: }\emph{How effective is the proposed approach in improving the relation between assertions and focal methods?} 

\textbf{RQ4 - Understandability: }\emph{What is the impact of the proposed approach on the understandability and maintainability of automatically generated tests?}

For \textbf{RQ1} we first propose different test case representations that reflect a structure that follows single-responsibility principle and allow us to identify focal methods. We plan to implement them in EvoSuite using the DynaMOSA algorithm~\cite{Fraser2011,panichella2018dynamosa}, discuss the changes needed in each representation (\eg variation in genetic operators, fitness function), and analyze the coverage and mutation score of each proposed structure.  

\textbf{RQ2} is focused on evaluating the proposed approach to structure test cases following the single-responsibility principle. For \textbf{RQ3}, we study the effects of our approach in assertion generation and their relation to the test focal method. To evaluate both, we plan to follow the methodology proposed in previous studies~\cite{panichella2018dynamosa,icsme2023} and evaluate the tests based on dynamic analysis~\cite{he2024empirical} and manual review.

Finally, \textbf{RQ4} is related to the perception of developers of the tests generated using the proposed approach. Thus, we plan to conduct an extensive user study to analyze the impact of using structured tests on understandability and maintainability. We aim to follow previous studies~\cite{shamshiri2018automatically,ceccato2015automatically} and gather actionable recommendations for future work. 

\section{Expected Contributions}
To the date of this proposal, we have not found a prior implementation that modifies the test case representation to generate structured tests based on the single-responsibility principle. By conducting this research, we will provide practitioners with a novel technique for structuring automatically generated tests implemented in EvoSuite. Our approach focuses on improving the understandability of the generated tests by explicitly identifying the relation between the test code and source code and validating it with developers, we aim to promote the use of generation tools in industrial development. The results of this research can also benefit researchers and tool developers by providing a defined structure to perform further quality analysis and reveal key factors for improving the use of generated tests allowing further future work on the topic.

\section*{Acknowledgments}
This research is supervised by Prof. Dr. Alexandre Bergel, in collaboration with Prof. Dr. Juan Pablo Sandoval Alcocer. I am grateful to Prof. Dr. Gordon Fraser for his kind reviews and valuable feedback on this proposal. 

\newpage
\bibliographystyle{IEEEtran}
\bibliography{references.bib}

\end{document}